\input{aipcheck}

\documentclass[
    ,final            
  ]
  {aipproc}

\layoutstyle{6x9}

\begin{document}

\title{Quantum Monte Carlo calculations of magnetic moments and M1 transitions in $A \le 9$ nuclei}

\author{S.~Pastore}{address={Physics Division, Argonne National Laboratory, Argonne, IL 60439, USA}
}
\author{Steven~C.~Pieper}{address={Physics Division, Argonne National Laboratory, Argonne, IL 60439, USA}
}
\author{R.~Schiavilla}{address={Theory Center, Jefferson Laboratory, Newport News, VA 23606, USA},
                       altaddress={Physics Department, Old Dominion University, Norfolk, VA 23529, USA}
}
\author{R.~B.~Wiringa}{  address={Physics Division, Argonne National Laboratory, Argonne, IL 60439, USA}
}

\begin{abstract}
We present Quantum Monte Carlo calculations of magnetic moments and M1 transitions
in $A\le 9$ nuclei which take into account contributions of two-body electromagnetic currents.
The Hamiltonian utilized to generate the nuclear wave functions includes the realistic
Argonne-{\it v}$_{18}$ two-nucleon and the Illinois-7 three-nucleon interactions.
The nuclear two-body electromagnetic currents are derived from a pionful chiral effective field theory
including up to one-loop corrections. These currents involve unknown Low Energy Constants
which have been fixed so as to reproduce a number of experimental data for the two- and three-nucleon
systems, such as $np$ phase shifts and deuteron, triton, and $^3$He magnetic moments.
This preliminary study shows that two-body contributions provide significant corrections
which are crucial to bring the theory in agreement with the experimental data in both
magnetic moments and M1 transitions.  
\end{abstract}

\maketitle

\section{Introduction}
\vspace{-0.05in}

Quantum Monte Carlo calculations of magnetic moments (m.m.'s) and M1 transitions in $A\le 7$ nuclei
have been presented in recent years by Marcucci {\it et al.} in Ref.~\cite{Marcucci08}.
In that work, the authors investigated the role played by the electromagnetic (EM) two-body
meson-exchange currents (MEC) derived from the realistic nucleon-nucleon (NN) interaction Argonne
{\it v}$_{18}$ (AV18) via current conservation~\cite{Marcucci08,Marcucci05}. It was found that MEC
contributions increase the $A=3,7$ isovector m.m.'s, as well as the $A=6,7$ M1 transitions
by $16\%$ and $17-34\%$, respectively, bringing them into very good agreement with experimental data. 
In this framework---also referred to as the Standard Nuclear Physics Approach (SNPA)---the isoscalar m.m.'s
of the $A=3$,$7$ nuclei are, however, underpredicted by a few percent (up to $10\%$ in $A=7$ systems)~\cite{Marcucci08}.
\begin{figure}
\includegraphics[height=.237\textheight]{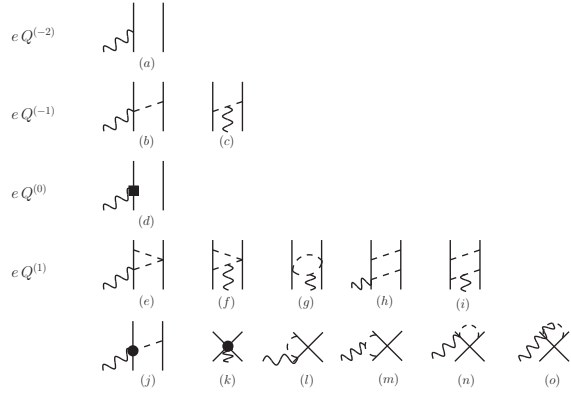}
\caption{Diagrams illustrating one- and two-body EM currents entering at LO ($e\, Q^{-2}$),
NLO ($e\, Q^{-1}$), N2LO  ($e\, Q^{\,0}$), and N3LO ($e\, Q^{\,1}$).  Nucleons, pions,
and photons are denoted by solid, dashed, and wavy lines, respectively.}
\label{fig:f1}
\end{figure}

Two-body EM current operators have been derived recently within pionful chiral
effective field ($\chi$EFT) formulations~\cite{Park96,Pastore08,Pastore09,Pastore11,Kolling09,Kolling11}.
The $\chi$EFT current operators are expanded in powers of pions' and nucleons'
momenta, and consist of long- and intermediate-range components which are described in
terms of one- and two-pion exchange contributions, as well as contact currents which
encode the short-range physics. These operators involve a number of Low Energy Constants (LECs)
which are then fixed to the experimental data. In hybrid calculations,
many-body operators derived from a $\chi$EFT framework, are utilized in transition matrix
elements in between nuclear wave functions (w.f.'s) obtained from Hamiltonians involving realistic
two- and three-body nuclear interactions. Intrinsic to this approach is a mismatch between
the short-range behavior of the NN potential and that of the current operators. 
Hybrid calculations of EM observables in the $A=2$--$4$ nuclei
(see for example Refs.~\cite{Park03,Song07,Song09,Lazauskas11,Girlanda10}) indicate
that this inconsistency is in most cases mitigated by the fitting procedure implemented to constrain
the LECs.

The analysis presented here has two objectives. The first one is
to extend the studies reported in Ref.~\cite{Marcucci08} to $A>7$ nuclei.
The second one is to investigate how the $\chi$EFT EM current operators of
Refs.~\cite{Pastore08,Pastore09}, {\it albeit} utilized within a hybrid context,
compare with the SNPA formulation of Ref.~\cite{Marcucci08}. In what follows we
briefly report on the Quantum Monte Carlo (QMC) techniques, and the $\chi$EFT EM
operators utilized in these calculations, and provide preliminary results obtained from them.
 
\section{QMC Method and the Nuclear Hamiltonian}
\vspace{-0.1in}

The EM transition matrix elements are evaluated in between w.f.'s which are solutions of the Schr\"odinger equation
\begin{equation}
 H | \Psi \rangle = E | \Psi \rangle \ .
\end{equation}
The nuclear Hamiltonian used in the calculations consists of a kinetic term plus
two- and three-body interaction terms, namely the AV18~\cite{AV18} and the
Illinois-7~\cite{IL7}, respectively:
\begin{equation}
 H = \sum_i K_i + \sum_{i<j} v_{ij} + \sum_{i<j<k} V_{ijk} \ .
\end{equation}
Nuclear w.f.'s are constructed in two steps. First, a trial variational Monte Carlo w.f. ($\Psi_T$),
which accounts for the effect of the nuclear interaction via the inclusion of correlation operators,
is generated by minimizing the energy expectation value with respect to a number of
variational parameters. The second step improves on $\Psi_T$ by eliminating
excited states contamination. This is accomplished in a Green's function Monte Carlo (GFMC)
calculation which propagates the Schr\"odinger equation in imaginary time ($\tau$). The propagated w.f.
$\Psi(\tau) = e^{-(H-E_0)\tau}\Psi_T$, for large values of $\tau$, converges to the
exact w.f. with eigenvalue $E_0$.  Ideally, the matrix elements should be evaluated in between
two propagated w.f.'s. In practice, we evaluate mixed estimates in which only one w.f. is propagated,
while the remaining one is replaced by $\Psi_T$. The calculation of diagonal and
off-diagonal matrix elements is discussed at length in Ref.~\cite{Pervin07} and references therein. 

The nuclear EM current operator---regardless of the formalism utilized to construct it---is also
expressed as an expansion in many-body operators.
The current utilized in this work accounts up to two-body effects, and is written as:
\begin{equation}
 {\bf j}({\bf q}) =  \sum_i {\bf j}_i({\bf q}) + \sum_{i<j} {\bf j}_{ij}({\bf q}) \ ,
\end{equation}
where ${\bf q}$ is the momentum associated with the external EM field. 

\section{EM currents and magnetic moments  in $\chi$EFT}
\vspace{-0.1in}

Currents from pionful $\chi$EFT including up to two-pion exchange contributions
were derived originally by Park, Min, and Rho in covariant perturbation theory~\cite{Park96}.
More recently, K\"olling and collaborators presented EM currents obtained within
the method of unitary transformations~\cite{Kolling09,Kolling11}. Here, we refer to the EM operators
constructed in Ref.~\cite{Pastore09}, in which time-ordered perturbation theory is
implemented to calculate the EM transition amplitudes. Differences between the models
mentioned above have been discussed in Refs.~\cite{Pastore09,Kolling11}, and will not
be addressed here. Instead, we limit ourselves to  briefly describing
the various contributions to the EM currents which have been utilized in the GFMC calculations,
and refer to Ref.~\cite{Pastore09} for the formal expressions of the operators. However,
we remark that we have revised our calculation~\cite{Girlanda12}
of the operators associated with diagrams (m) and (o) of Fig.~\ref{fig:f1}. 

The $\chi$EFT EM current operators are diagrammatically represented in Fig.~\ref{fig:f1}.
They are expressed as an expansion in $Q$, {\it i.e.},
the low-momentum scale.  Referring to Fig.~\ref{fig:f1}, the leading-order (LO) term is counted
as $e\,Q^{-2}$ ($e$ is the electric charge), and consists of the single-nucleon
convection and spin-magnetization currents. The NLO term (of order $e\, Q^{-1}$)
involves seagull and in-flight contributions associated with one-pion exchange,
and the N2LO term (of order $e\, Q^0$) represents the $(Q/m_N)^2$ relativistic
correction to the LO one-body current ($m_N$ denotes the
nucleon mass). At N3LO ($e\, Q$) we include one-loop contributions of diagrams
(e)--(i) and (l)--(o), as well as the tree-level current involving a
$\gamma \pi NN$ vertex of order $e\, Q^2$---of diagram (j), and the contact currents of
diagram (k). The two-body operators have a power-law behavior at large momenta, therefore
a regularization procedure is implemented via the introduction of cutoff function of the form 
$exp(-Q^4/\Lambda^4)$~\cite{Girlanda12}, where $\Lambda=600$ MeV.

The contact currents of diagram (k) involve both minimal and non-minimal
LECs. Minimal LECs enter the $\chi$EFT contact NN interaction
at order $Q^2$, and can be taken from fits to the NN scattering
data. We use the values obtained from the analysis of Refs.~\cite{Entem03,Machleidt11}, with cutoff
$\Lambda=600$ MeV.    
Non-minimal LECs entering the contact and tree-level currents at N3LO---diagrams (j)
and (k), respectively---need to be fixed to EM observables. The contact currents involve
two LECs, multiplying an isoscalar and an isovector operator, respectively. 
There are three LECs entering the tree-level current of diagram (j). Two of them
multiply isovector structures and they saturate the  $\Delta$-resonance
excitation current, while the third one is associated with
an isoscalar operator and saturates the $\rho\pi\gamma$ transition current~\cite{Park96}.
We exploit the  $\Delta$-resonance saturation mechanism, thus reducing the number of
unknown LECs to three. We fix the two isoscalar LECs so as to reproduce the deuteron
and the isoscalar combination of the trinucleon m.m.'s, while the isovector
LEC is obtained from fits to the isovector combination of the $A=3$ nuclei m.m.'s. This choice
provides us with the most natural LECs~\cite{Girlanda12}.

\begin{figure}
\begin{minipage}[t]{0.40\textheight}
\includegraphics[height=.368\textheight,angle=270,keepaspectratio=true]{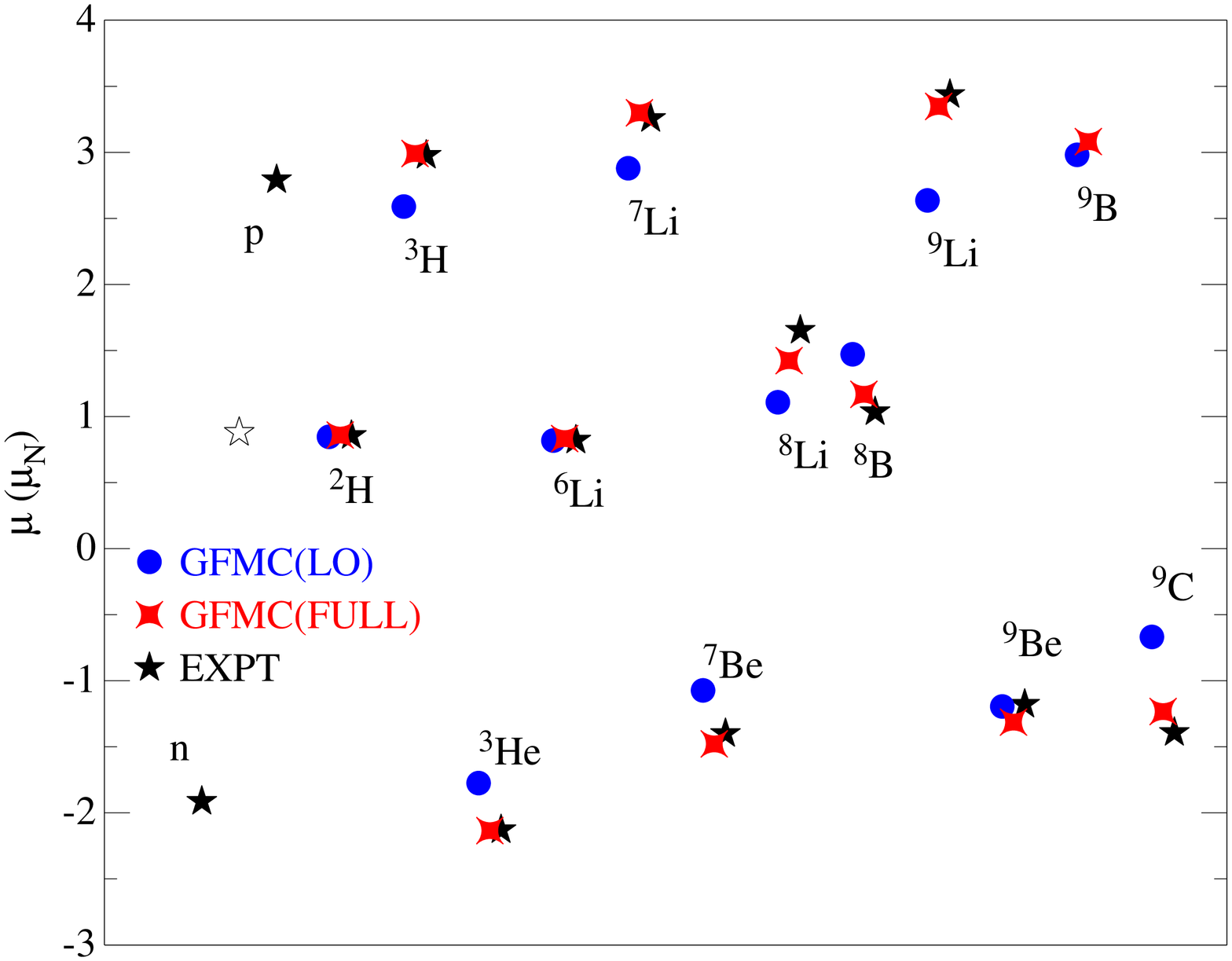}
\end{minipage}
\begin{minipage}[t]{0.25\textheight}
\includegraphics[height=.268\textheight,angle=270,keepaspectratio=true]{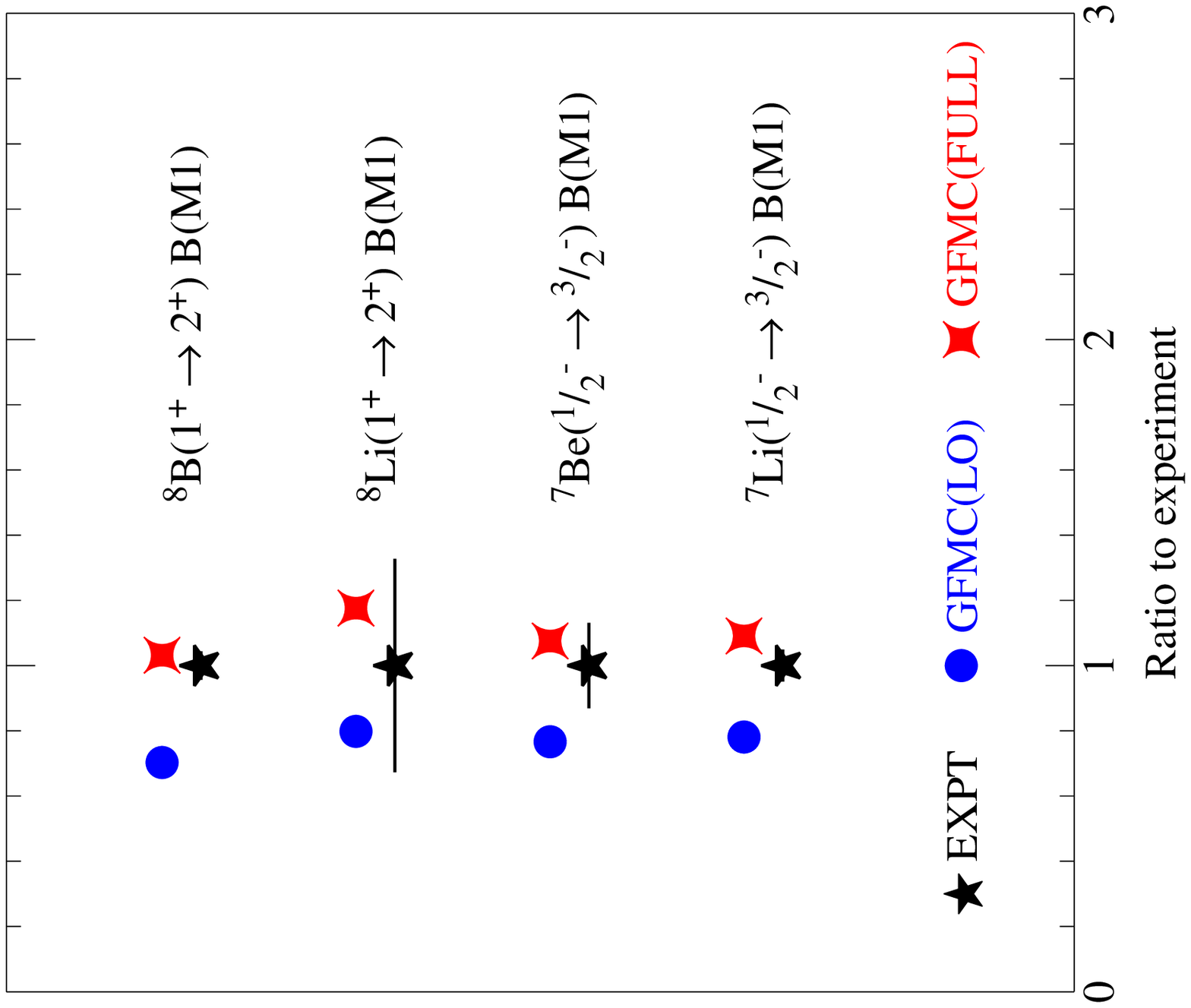}
\caption{  {\bf Left}: Magnetic moments in nuclear magnetons for $A\leq9$ nuclei. 
           Black stars indicate
           the experimental values~\cite{Tilley02,Tilley04}, while blue dots (red diamonds)
           represent preliminary GFMC calculations which include the LO one-body EM current
           (full $\chi$EFT current up to N3LO). Predictions are for nuclei with $A>3$. {\bf Right}:
           Transition widths normalized to the experimental values~\cite{Tilley02,Tilley04}
           for $A=7$--$8$ nuclei, notation as in left panel.
           }
\label{fig:f2}
\end{minipage}
\end{figure}

\section{Results}
\vspace{-0.05in}

The preliminary results for the m.m.'s of $A\leq 9$ nuclei are summarized in the left panel of Fig.~\ref{fig:f2}.
In this figure, black stars represent the experimental data~\cite{Tilley02,Tilley04}---there are no data
for the m.m. of $^9$B. For completeness, we show also the
experimental values for the proton and neutron m.m.'s, as well as their sum, which corresponds to the m.m.
of an S-wave deuteron. The experimental values of the $A=2$--$3$ m.m.'s have been utilized to fix the LECs,
therefore predictions are for $A>3$ nuclei. The blue dots labeled as GFMC(LO) represent theoretical
predictions obtained with the standard one-nucleon EM current entering at LO---diagram a) of Fig.~\ref{fig:f1}.
The GFMC(LO) results reproduce the bulk properties of the m.m.'s
of the light nuclei considered here. In particular, we can recognize three classes of nuclei, that is
nuclei whose m.m.'s  are driven by an unpaired valence proton, or neutron,
or `deuteron cluster' inside the nucleus. 
Predictions which include all
the contributions to the N3LO $\chi$EFT EM currents illustrated in Fig.~\ref{fig:f1} are represented by the red
diamonds of Fig.~\ref{fig:f2}, labeled GFMC(FULL). In most of the cases considered here,
the predicted m.m.'s  are closer to the experimental data when the corrections entering at NLO and following orders are
added to the LO one-body EM operator. Notable are the cases associated with the $A=9$ and $T=3/2$
nuclei, in which these corrections are found to provide up to $\sim 40 \%$ of the total predictions.
We also tested how the SNPA currents perform for these EM observables
and found that the hybrid $\chi$EFT formulation provides us with improved values for the isoscalar
m.m.'s  of the nuclei considered here. 

In the right panel of Fig.~\ref{fig:f2}, we show the preliminary result for M1 transitions in $A\leq 8$ nuclei. Here,
we show the ratios to the experimental values of the widths~\cite{Tilley02,Tilley04}. The latter
are represented with the black stars along with the associated experimental error bars,
while the GFMC(LO) and GFMC(FULL) predictions are again represented by blue dots and red diamonds, respectively.
Also for these EM observables, predictions which account for the complete N3LO operator
are closer to the experimental values, but for the transition in $^8$Li, for which the experimental
error is large, we cannot determine whether the GFMC(FULL) prediction is a better one.
The study presented here is at its first stage. A manuscript with a detailed presentation of this work
is in preparation~\cite{Pastore12}.


\begin{theacknowledgments}
\vspace{-0.05in}
The many-body calculations were performed on the parallel computers of the
Laboratory Computing Resource Center, Argonne National Laboratory.
This work is supported by the U. S. Department of Energy,
Office of Nuclear Physics, under contracts No. DE-AC02-06CH11357
and DE-AC05-06OR23177 and under SciDAC grant No. DE-FC02-07ER41457.
\end{theacknowledgments}

\bibliographystyle{aipproc} 
\bibliography{420_Pastore}

\hyphenation{Post-Script Sprin-ger}
\begin{thebibliography}{22}
\expandafter\ifx\csname natexlab\endcsname\relax\def\natexlab#1{#1}\fi
\providecommand{\enquote}[1]{``#1''}
\expandafter\ifx\csname url\endcsname\relax
  \def\url#1{\texttt{#1}}\fi
\expandafter\ifx\csname urlprefix\endcsname\relax\def\urlprefix{URL }\fi
\providecommand{\eprint}[2][]{\url{#2}}

\bibitem[Marcucci et~al.(2008)]{Marcucci08}
L.~E. Marcucci, M.~Pervin, S.~C. Pieper, R.~Schiavilla, and R.~B. Wiringa,
  \emph{Phys. Rev. C} \textbf{78}, 065501 (2008).

\bibitem[Marcucci et~al.(2005)]{Marcucci05}
L.~E. Marcucci, M.~Viviani, R.~Schiavilla, A.~Kievsky, and S.~Rosati,
  \emph{Phys. Rev. C} \textbf{72}, 014001 (2005).

\bibitem[Park et~al.(1996)]{Park96}
T.-S. Park, D.-P. Min, and M.~Rho, \emph{Nuclear Physics A} \textbf{596}, 515
  (1996).

\bibitem[Pastore et~al.(2008)]{Pastore08}
S.~Pastore, R.~Schiavilla, and J.~L. Goity, \emph{Phys. Rev. C} \textbf{78},
  064002 (2008).

\bibitem[Pastore et~al.(2009)]{Pastore09}
S.~Pastore, L.~Girlanda, R.~Schiavilla, M.~Viviani, and R.~B. Wiringa,
  \emph{Phys. Rev. C} \textbf{80}, 034004 (2009).

\bibitem[Pastore et~al.(2011)]{Pastore11}
S.~Pastore, L.~Girlanda, R.~Schiavilla, and M.~Viviani, \emph{Phys. Rev. C}
  \textbf{84}, 024001 (2011).

\bibitem[K\"olling et~al.(2009)]{Kolling09}
S.~K\"olling, E.~Epelbaum, H.~Krebs, and U.~G. Mei\ss{}ner, \emph{Phys. Rev. C}
  \textbf{80}, 045502 (2009).

\bibitem[K\"olling et~al.(2011)]{Kolling11}
S.~K\"olling, E.~Epelbaum, H.~Krebs, and U.-G. Mei\ss{}ner, \emph{Phys. Rev. C}
  \textbf{84}, 054008 (2011).

\bibitem[Park et~al.(2003)]{Park03}
T.-S. Park, L.~E. Marcucci, R.~Schiavilla, M.~Viviani, A.~Kievsky, S.~Rosati,
  K.~Kubodera, D.-P. Min, and M.~Rho, \emph{Phys. Rev. C} \textbf{67}, 055206
  (2003).

\bibitem[Song et~al.(2007)]{Song07}
Y.-H. Song, R.~Lazauskas, T.-S. Park, and D.-P. Min, \emph{Physics Letters B}
  \textbf{656}, 174 (2007).

\bibitem[Song et~al.(2009)]{Song09}
Y.-H. Song, R.~Lazauskas, and T.-S. Park, \emph{Phys. Rev. C} \textbf{79},
  064002 (2009).

\bibitem[Lazauskas et~al.(2011)]{Lazauskas11}
R.~Lazauskas, Y.-H. Song, and T.-S. Park, \emph{Phys. Rev. C} \textbf{83},
  034006 (2011).

\bibitem[Girlanda et~al.(2010)]{Girlanda10}
L.~Girlanda, A.~Kievsky, L.~E. Marcucci, S.~Pastore, R.~Schiavilla, and
  M.~Viviani, \emph{Phys. Rev. Lett.} \textbf{105}, 232502 (2010).

\bibitem[Wiringa et~al.(1995)]{AV18}
R.~B. Wiringa, V.~G.~J. Stoks, and R.~Schiavilla, \emph{Phys. Rev. C}
  \textbf{51}, 38 (1995).

\bibitem[Pieper(2008)]{IL7}
S.~C. Pieper, \emph{AIP Conference Proceedings} \textbf{1011}, 143 (2008).

\bibitem[Pervin et~al.(2007)]{Pervin07}
M.~Pervin, S.~C. Pieper, and R.~B. Wiringa, \emph{Phys. Rev. C} \textbf{76},
  064319 (2007).

\bibitem[Girlanda et~al.(in preparation)]{Girlanda12}
L.~Girlanda, S.~Pastore, R.~Schiavilla, and M.~Viviani  (in preparation).

\bibitem[Entem and Machleidt(2003)]{Entem03}
D.~R. Entem, and R.~Machleidt, \emph{Phys. Rev. C} \textbf{68}, 041001 (2003).

\bibitem[Machleidt and Entem(2011)]{Machleidt11}
R.~Machleidt, and D.~Entem, \emph{Physics Reports} \textbf{503}, 1 (2011).

\bibitem[Tilley et~al.(2002)]{Tilley02}
D.~Tilley, C.~Cheves, J.~Godwin, G.~Hale, H.~Hofmann, J.~Kelley, C.~Sheu, and
  H.~Weller, \emph{Nuclear Physics A} \textbf{708}, 3 (2002).

\bibitem[Tilley et~al.(2004)]{Tilley04}
D.~Tilley, J.~Kelley, J.~Godwin, D.~Millener, J.~Purcell, C.~Sheu, and
  H.~Weller, \emph{Nuclear Physics A} \textbf{745}, 155 (2004).

\bibitem[Pastore et~al.(in preparation)]{Pastore12}
S.~Pastore, S.~C. Pieper, R.~Schiavilla, and R.~B. Wiringa  (in preparation).

\end{thebibliography}
%


\end{document}